\documentclass[12pt, fleqn, a4paper]{article} 
 \usepackage{graphicx, subcaption}
 \usepackage{ifthen}
 \usepackage{color}
 \usepackage{clshan-math}
 \newcommand{\Gau}   {{\rm Gau}}
 \newcommand{\vesc}  {v_{\rm esc}}
 \newcommand{\vSunG} {v_{\rm \odot, G}}
 \definecolor{green} {rgb} {0   , 0.5 , 0   } 
 \definecolor{white} {rgb} {1   , 1   , 1   } 
 \def \PeriodAa     {0      -- 365}
 \def \PlotNumberAa {00000}
 \def \PeriodCa     {19.49  --  79.49}
 \def \PeriodCb     {110.74 -- 170.74}
 \def \PeriodCc     {201.99 -- 261.99}
 \def \PeriodCd     {293.24 -- 353.24}
 \def \PlotNumberCa {04949}
 \def \PlotNumberCb {14074}
 \def \PlotNumberCc {23199}
 \def \PlotNumberCd {32324}
\newcommand{\InsertPlotPphithetaS} [1] {%
 \begin{minipage} {\Subplotwidth}
  {\begin{center}
    \ifthenelse {\equal {\vzeroValue} {220}}
                {\includegraphics [width = \Subplotwidth]
                                  {PoN_phi_theta-Eq-500-#1}}
                {\ifthenelse {\equal {\vSunGValue} {220}}
                             {\includegraphics [width = \Subplotwidth]
                                               {PoN_phi_theta-Eq-500-#1-\vzeroValue}}
                             {\includegraphics [width = \Subplotwidth]
                                               {PoN_phi_theta-Eq-500-#1-\vSunGValue-\vzeroValue}}}
   \end{center}}
 \end{minipage}
}
\newcommand{\InsertPlotPeriodS} [1] {%
 \begin{minipage} {\Subplotwidth}
  {\begin{center}
    {\footnotesize #1}
   \end{center}}
 \end{minipage}
}
\newcommand{\InsertFBoxPlotPphithetavzero} [2] [{green} {white}] {
 \fcolorbox #1
  {\begin{subfigure} [c] {17 cm}
    {\begin{center}
     \raisebox{-2.5 cm} [0 pt] [0 pt] {\InsertPlotPphithetaS {\PlotNumberAa}}\hspace{0.2 cm}
     \InsertPlotPphithetaS {\PlotNumberCa}\hspace{0.2 cm}
     \InsertPlotPphithetaS {\PlotNumberCb}
     \\ \vspace{0.1 cm}
     \raisebox{-2.5 cm} [0 pt] [0 pt] {\InsertPlotPeriodS    {\PeriodAa\ day}}\hspace{0.2 cm}
     \InsertPlotPeriodS    {\PeriodCa\ day}\hspace{0.2 cm}
     \InsertPlotPeriodS    {\PeriodCb\ day}
     \\ \vspace{0.1 cm}
     \begin{minipage} {\Subplotwidth}
      {\begin{center}
        ~
       \end{center}}
     \end{minipage}
     \hspace{0.2 cm}
     \InsertPlotPphithetaS {\PlotNumberCc}\hspace{0.2 cm}
     \InsertPlotPphithetaS {\PlotNumberCd}
     \\ \vspace{0.1 cm}
     \begin{minipage} {\Subplotwidth}
      {\begin{center}
        ~
       \end{center}}
     \end{minipage}
     \hspace{0.2 cm}
     \InsertPlotPeriodS    {\PeriodCc\ day}\hspace{0.2 cm}
     \InsertPlotPeriodS    {\PeriodCd\ day}
     \\ \vspace{0.1 cm}%
     \caption{#2}
     \end{center}}
   \end{subfigure}}
}
\newcommand{\InsertFigurePphithetavzero} [2]
 [As in Figs.~\ref{fig:PoN_phi_theta-500-180},
  except that
  $\vSunG = v_0 = \vzeroValue$ km/s
  (a)
  as well as
  $\vSunG = 220$ km/s and $v_0 = \vzeroValue$ km/s
  (b)
  have been simulated.%
  ]
{
 \def \vzeroValue {#2}
\begin{figure} [t!]
\begin{center}
 \def \Subplotwidth {5.3 cm}
 \def \vSunGValue   {\vzeroValue}
 \InsertFBoxPlotPphithetavzero
  {$\vSunG = v_0 = \vzeroValue$ km/s}
 \\ \vspace{0.1 cm}
 \def \vSunGValue   {220}
 \InsertFBoxPlotPphithetavzero
  {$\vSunG = 220$ km/s and $v_0 = \vzeroValue$ km/s}
 \\ \vspace{-0.25 cm}
\end{center}
\caption{
 #1
}
\label{fig:PoN_phi_theta-500-\vzeroValue}
\end{figure}
}
\begin{document}
\thispagestyle{empty}
\begin{flushright}
 November 2020
\end{flushright}
\begin{center}
{\large\bf
 Dependence of
 the WIMP Angular Kinetic--Energy Distribution on \\ \vspace{0.15 cm}
 the Solar Galactic Orbital Velocity}             \\
\vspace*{0.7 cm}
 {\sc Chung-Lin Shan}                             \\
\vspace{0.5 cm}
 {\it Preparatory Office of
      the Supporting Center for
      Taiwan Independent Researchers              \\ \vspace{0.05 cm}
      P.O.BOX 21 National Chiao Tung University,
      Hsinchu City 30099, Taiwan, R.O.C.}         \\~\\~\\
 {\it E-mail:} {\tt clshan@tir.tw}
\end{center}
\vspace{2 cm}
\begin{abstract}

 In this article,
 we study in a little more detail
 the angular kinetic--energy distribution of
 halo Weakly Interacting Massive Particles (WIMPs)
 and consider
 two simple modifications with
 the Solar Galactic orbital velocity
 in our Monte Carlo simulations of
 the 3-dimensional WIMP velocity
 as the first trial
 of future investigations on
 distinguishing models of
 the Galactic structure of Dark Matter particles
 by using directional direct detection data.

\end{abstract}
\clearpage
\section{Introduction}

 Directional direct Dark Matter (DM) detection experiments
 aim to measure 3-dimensional information
 (recoil tracks and/or head--tail senses) of
 target nuclei
 scattered by
 halo Weakly Interacting Massive Particles (WIMPs)
 \cite{Ahlen09, Mayet16}.
 This could provide a practical way
 for discriminating WIMP singles from
 neutrino--nucleus scattering backgrounds
 (the so--called ``neutrino floor'')
 \cite{OHare15b, OHare17, OHare20}.

 In our earlier works
 \cite{DMDDD-N, DMDDD-P},
 we investigated
 the angular distributions of
 the direction (flux)
 as well as
 the accumulated and the average kinetic energies of
 the Monte Carlo--simulated 3-dimensional WIMP velocity
 in different celestial coordinate systems.
 While,
 in the Equatorial coordinate system,
 the anisotropy
 and
 the directionality (clockwise rotated annual modulation) of
 the distribution patterns of
 the 3-D WIMP direction/kinetic energy
 could be clearly observed,
 two significant characteristics of
 the patterns of
 the ``average'' WIMP kinetic energy
 have also been discovered.

 Firstly,
 the (main) hot--points of
 the angular distribution patterns of
 the {\em average} WIMP kinetic energy
 appearing approximately in the straight area
 between the longitude of 90$^{\circ}$W and 0$^{\circ}$
 and
 below the latitude of 15$^{\circ}$S
 could not coincide with
 the hot--points of
 the patterns of
 the WIMP flux and
 the accumulated kinetic energy
 spreading obliquely
 from the center to the southwest part of the sky
 \cite{DMDDD-P}.
 Secondly,
 two extra hot--points
 close to the center and
 at the southwestern corner of the sky
 could also be observed
 in the distribution patterns of
 the average WIMP kinetic energy.
 Accompanied with the annual modulation of
 the distribution patterns,
 these two hot--points would even show
 a clearly seasonal appearance
 \cite{DMDDD-P}.

 So far
 we have
 theoretical explanations for
 neither the pattern difference
 nor the (seasonal) appearance of the extra hot--points.
 Nevertheless,
 in order to understand
 in more detail
 their contributions to
 our future study on
 the Galactic structure of DM particles
 by using directional direct detection data,
 in this article,
 we investigate
 the dependence of
 the angular distribution of
 the WIMP average kinetic energy
 on the Solar Galactic orbital velocity.

 In the next section,
 we will consider two modifications of
 the standard Galactic halo model
 used in our simulations
 and discuss the corresponding distribution patterns of
 the 3-D WIMP average kinetic energy.
 Then
 we conclude in Sec.~3.

\section{Two modifications of the standard Galactic halo model}

 In our Monte Carlo simulations of
 the 3-dimensional information on
 the WIMP velocity,
 the simple Maxwellian velocity distribution function
 truncated at the Galactic escape velocity
 has been adopted
 for the radial component
 in the Galactic coordinate system
 \cite{SUSYDM96}:
\beq
     f_{\rm G, r}(v)
  =  f_{1, \Gau}(v)
  =  \bbrac{  \afrac{\sqrt{\pi}}{4} \erf\afrac{\vesc}{v_0}
            - \afrac{\vesc}{2 v_0}  e^{-\vesc^2 / v_0^2}   }^{-1}
     \afrac{v^2}{v_0^3}
     e^{-v^2 / v_0^2}
\~,
\label{eqn:f1v_Gau_vesc}
\eeq
 for $v \le \vesc$,
 and $f_{\rm G, r}(v > \vesc) = 0$.
 Here
 $v_0$
 is the Solar orbital velocity around the Galactic center,
 $\vesc$
 is the escape velocity from our Galaxy
 at the position of the Solar system
 and has been set as 550 km/s in our simulations
 \cite{RPP20AP}.
 Moreover,
 for
 the angular component of the \mbox{3-D} WIMP velocity,
 a simple isotropic distribution
 in the Galactic coordinate system
 has been considered
 \cite{DMDDD-N}.

 Meanwhile,
 each generated 3-D WIMP velocity
 (including the measuring time)
 will be transformed
 from the Galactic frame
 through the Ecliptic and the Equatorial frames
 to the laboratory frame
 (see Ref.~\cite{DMDDD-N} for details).
 For the transformation
 between the Galactic and the Ecliptic coordinate systems,
 the moving velocity of the Solar system in the Galaxy
 $\vSunG \simeq 220$ km/s
 has been adopted%
\footnote{
 Note that,
 although
 $v_0$
 in Eq.~(\ref{eqn:f1v_Gau_vesc})
 should theoretically be the same quantity as $\vSunG$,
 in this work
 we handle
 $\vSunG$ as the astronomical measurement
 but $v_0$ is purely a tunable parameter
 in the simple Maxwellian halo model.
}.

 In this article,
 as the first trial
 of future investigations on
 distinguishing models of
 the Galactic structure of Dark Matter particles
 by using directional direct detection data,
 we consider
 two simple scenarios with
 the Solar Galactic orbital velocity
 in our simulations
 and discuss the effects on
 the angular distribution patterns of
 the average kinetic energy of
 halo WIMPs
 in the Equatorial coordinate systems.

\subsection{\boldmath
            Varying $\vSunG$ and $v_0$ simultaneously}

 We consider at first
 the modifications of
 the Galactic orbital velocity of the Solar system $\vSunG$
 and,
 correspondingly,
 the value of the parameter $v_0$
 appearing in
 our generating distribution (\ref{eqn:f1v_Gau_vesc}).

 \setlength{\fboxrule}{1.5 pt}
 \InsertFigurePphithetavzero
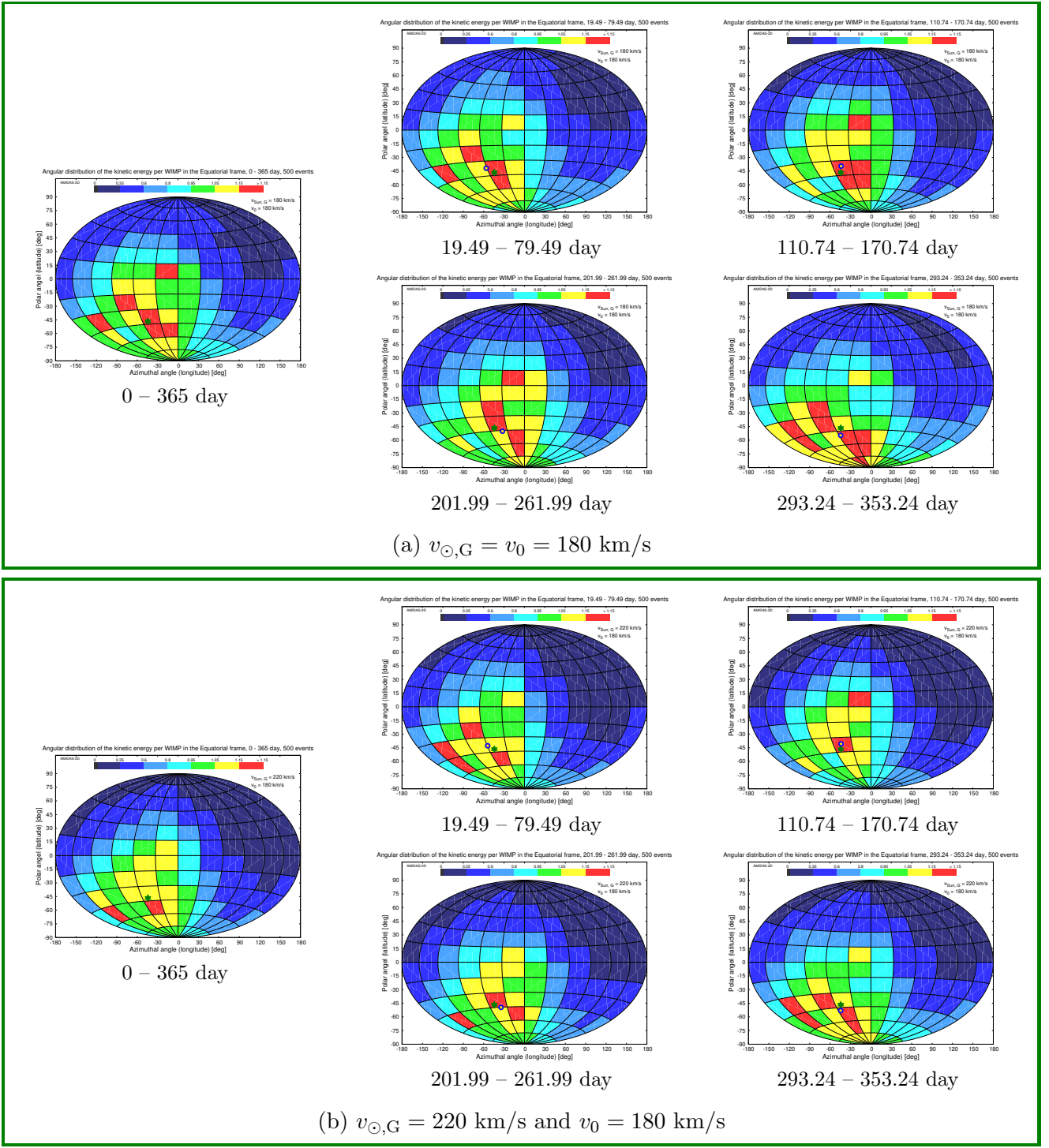
  [The angular distributions of
   the WIMP average kinetic energy
   in the Equatorial coordinate system
   with 500 total events on average
   in one entire year
   and
   in each 60-day observation period of
   four {\em advanced} seasons
   \cite{DMDDD-N}.
   (a)
   $\vSunG = v_0 = \vzeroValue$ km/s;
   (b)
   $\vSunG = 220$ km/s and \mbox{$v_0 = \vzeroValue$ km/s}.
   See the text for further details.
   ]
   {180}
 \InsertFigurePphithetavzero {200}
 \def \vzeroValue {220}
\begin{figure} [t!]
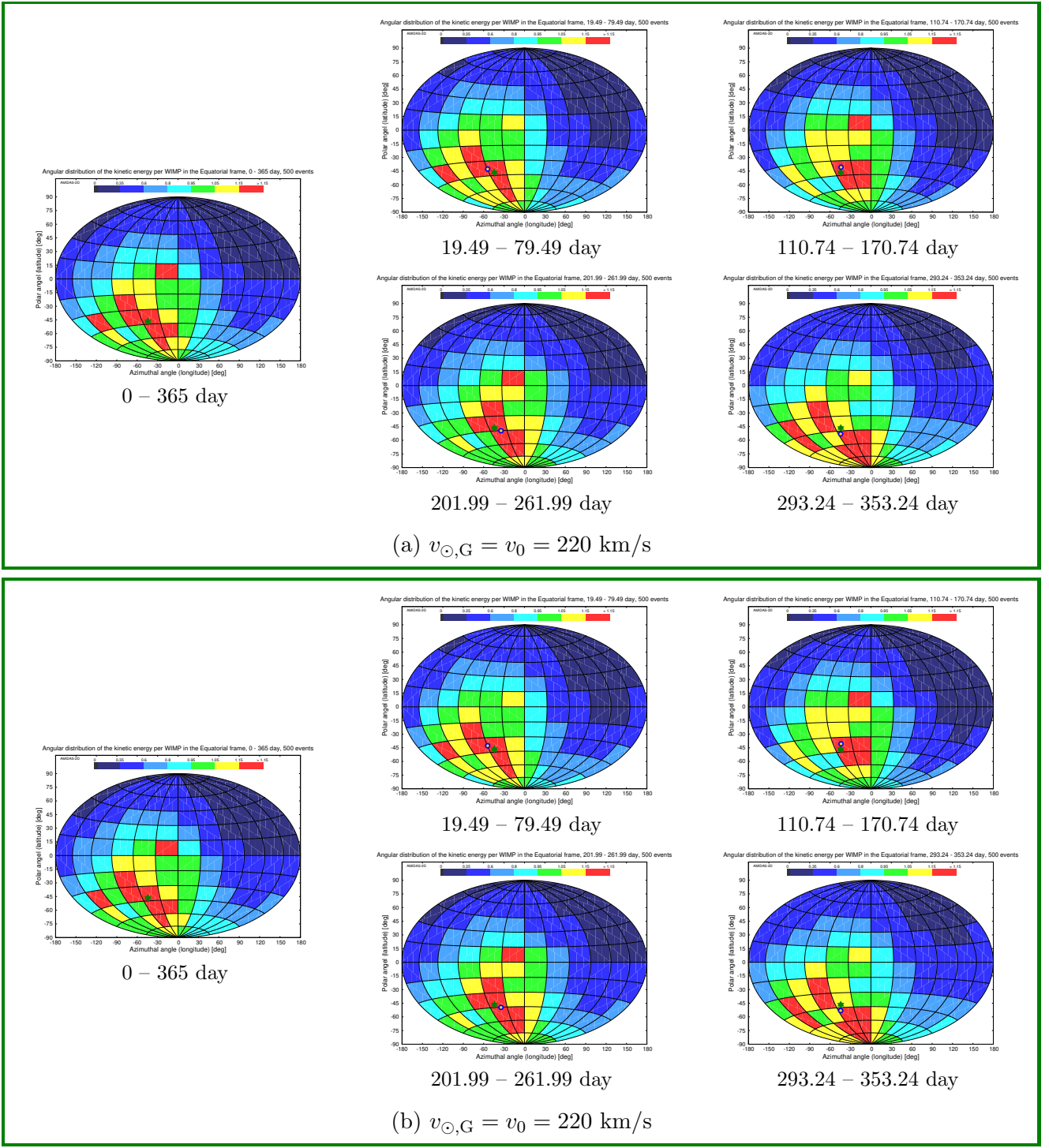

\begin{center}
 \def \Subplotwidth {5.3 cm}
 \def \vSunGValue   {\vzeroValue}
 \InsertFBoxPlotPphithetavzero
  {$\vSunG = v_0 = \vzeroValue$ km/s}
 \\ \vspace{0.1 cm}
 \def \vSunGValue   {220}
 \InsertFBoxPlotPphithetavzero
  {$\vSunG = v_0 = \vzeroValue$ km/s}
 \\ \vspace{-0.25 cm}
\end{center}
\caption{
 As in Figs.~\ref{fig:PoN_phi_theta-500-180},
 except that
 the results with
 $\vSunG = v_0 = \vzeroValue$ km/s
 have been given
 in both of the upper and lower frames
 as the comparison standard
 (figures from Ref.~\cite{DMDDD-P}).
}
\label{fig:PoN_phi_theta-500-\vzeroValue}
\end{figure}
 \InsertFigurePphithetavzero {240}
 \InsertFigurePphithetavzero {260}

 In the upper (a) frame
 of Figs.~\ref{fig:PoN_phi_theta-500-180}
 to \ref{fig:PoN_phi_theta-500-260},
 we show
 the angular distributions of
 the 3-D WIMP average kinetic energy
 in the Equatorial coordinate system
 with
 500 total events on average
 (Poisson--distributed)
 in one experiment
 in one entire year
 and
 in each 60-day observation period of
 four {\em advanced} seasons
 \cite{DMDDD-N},
 binned into 12 $\times$ 12 bins
 for the longitude and latitude directions,
 respectively.
 5,000 experiments
 with
 five values of
 $\vSunG = v_0 = 180$, 200, 220, 240, and \mbox{260 km/s}
 have been simulated.
 The comparison standard
 with $\vSunG = v_0 = 220$ km/s
 is given in Figs.~\ref{fig:PoN_phi_theta-500-220}(a).

 In each plot,
 the dark--green star indicates
 the theoretical main direction of
 the WIMP wind
 (the opposite direction of
  the Galactic movement of
  the Solar system)
 in the Equatorial coordinate system:
 42.00$^{\circ}$S, 50.70$^{\circ}$W.
 Additionally,
 in the plot
 simulated for each season,
 we also put
 a blue--yellow point
 to indicate
 the opposite direction of
 the Earth's velocity relative to the Dark Matter halo
 on the central date of each observation period%
\footnote{
 Note that
 the direction (the right ascensions and the declinations) of
 the Earth's velocity relative to the DM halo
 depends on the Solar orbital velocity $\vSunG$.
 Detailed calculations can be found
 in Appendix A of Ref.~\cite{DMDDD-N}.
}.
 Moreover,
 the horizontal color bar on the top of each plot
 indicates
 the mean value of the average kinetic energy
 (averaged over all simulated experiments)
 in each angular bin
 in unit of the all--sky average value%
\footnote{
 Note that
 the all--sky average value of
 the average WIMP kinetic energy
 depends (strongly) on
 the Earth's relative velocity to the DM halo
 and in turn
 the Solar orbital velocity $\vSunG$.
 Hence,
 the distribution pattern in each plot
 is {\em not} normalized by the same standard,
 but show only the relative strength
 in each single case
 with the specified $\vSunG$ and/or $v_0$ value
 and the observation period.
}.

 By comparing our simulation results,
 one would find that
 the difference
 between the distribution patterns
 (with each pair of the simultaneously varied $\vSunG$ and $v_0$)
 is not significant.
 This would indicate that,
 as long as
 we adopt the (actual) Solar orbital velocity $\vSunG$
 for the parameter $v_0$
 in the generating velocity distribution
 (\ref{eqn:f1v_Gau_vesc}),
 the angular distribution of
 the average WIMP kinetic energy
 would depend only slightly on
 $\vSunG$
 and
 could be a characteristic of
 the considered simple Maxwellian halo model.
 Certainly,
 with quantitative information about
 the (average) WIMP kinetic energy,
 the angular (average--)kinetic--energy distribution
 could provide a more precise value of $\vSunG$
 as well as
 the reliability of
 our halo model.

\subsection{\boldmath
            Fixing $\vSunG$ but varying $v_0$}

 Now,
 we consider a practical scenario
 in which
 the Solar orbital velocity
 is fixed as \mbox{$\vSunG = 220$ km/s},
 while
 the parameter $v_0$
 in our generating velocity distribution (\ref{eqn:f1v_Gau_vesc})
 varies alone
 from 180 km/s to 260 km/s,
 as a pre--study of
 the sensitivity of
 the angular distribution of
 the average WIMP kinetic energy
 for distinguishing predictions of different halo models.

 We show
 in the lower (b) frame
 of Figs.~\ref{fig:PoN_phi_theta-500-180}
 to \ref{fig:PoN_phi_theta-500-260}
 the angular distributions of
 the 3-D WIMP average kinetic energy
 with fixed $\vSunG = 220$ km/s
 and
 five values of
 $v_0 = 180$, 200, 220, 240, and 260 km/s,
 respectively,
 in the Equatorial coordinate system.
 Again,
 500 total events on average
 in one entire year
 and
 in each 60-day observation period of
 four advanced seasons
 has been considered
 and
 the comparison standard
 with $\vSunG = v_0 = 220$ km/s
 is given in Figs.~\ref{fig:PoN_phi_theta-500-220}(b).

 In contrast to
 the small difference
 between the distribution patterns
 with the simultaneously varied $\vSunG$ and $v_0$
 shown previously,
 the angular distributions
 with the fixed $\vSunG$ and the varied $v_0$
 show a significantly clear variation:
 the most energetic directions of incident WIMPs
 (the red hot--points)
 spread with the increased $v_0$,
 even though
 the all--sky average value
 in each plot
 increases already strongly with the increased $v_0$.
 This indicates that,
 even without quantitative information about
 the (average) WIMP kinetic energy,
 the angular distribution pattern of
 the average WIMP kinetic energy
 should already be sensitive to
 the value of
 the parameter $v_0$
 in the generating WIMP velocity distribution
 (\ref{eqn:f1v_Gau_vesc})
 and,
 in turn,
 would be useful
 for distinguishing models of
 the structure of halo Dark Matter.

\section{Summary}

 In this article,
 we investigated
 the dependence of
 the angular distribution of
 the WIMP average kinetic energy
 on the Galactic orbital velocity of the Solar system,
 as the first trial
 of future investigations on
 distinguishing models of
 the Galactic structure of Dark Matter particles
 by using directional direct detection data.

 Two simple scenarios with
 the Solar Galactic orbital velocity
 have been considered.
 At first
 we modified
 the Solar orbital velocity
 in the transformation between
 the Galactic and the Ecliptic coordinate systems
 and
 the parameter $v_0$
 appearing in
 our generating distribution
 for the radial component  of
 the 3-D WIMP velocity
 in the Galactic coordinate system
 simultaneously.
 Our simulations indicate that
 the simulated angular distribution of
 the average WIMP kinetic energy
 would depend only slightly on
 the Solar orbital velocity
 and
 could be a characteristic of
 the considered (simple Maxwellian) halo model.

 On the other hand,
 we considered also the scenario
 in which
 the Solar orbital velocity
 is fixed,
 while
 the parameter $v_0$
 in our generating distribution
 for
 the 3-D WIMP Galactic velocity
 varies.
 The spread of
 the most energetic directions of incident WIMPs
 in the simulated angular distributions
 with the increased $v_0$
 indicates that,
 even without quantitative information about
 the (average) WIMP kinetic energy,
 the angular average--kinetic--energy distribution
 should already be sensitive to
 the value of
 the parameter $v_0$
 and
 would be useful
 for distinguishing models of
 the structure of halo Dark Matter.

\subsubsection*{Acknowledgments}

 The author would like to thank
 the friendly hospitality of
 the Institute of Physics
 at the National Chiao Tung University
 during the finalization of this article.

%
%
%

%
%

%
%
%
\end{document}